\documentclass[amsfonts, amssymb, amsmath, reprint, showkeys, twoside, prb, floatfix]{revtex4-2}
\usepackage[english]{babel}

\usepackage[utf8]{inputenc}
\usepackage[T1]{fontenc}

\usepackage{amsthm}
\usepackage{mathtools}
\usepackage{units}

\usepackage{xcolor}
\usepackage[pdftex]{graphicx}

\usepackage{hyperref}
\hypersetup{colorlinks=true, citecolor=blue, urlcolor=blue, linkcolor=blue}

\usepackage{multirow}

\newcommand{\bra}[1]{\langle#1\rvert} 
\newcommand{\ket}[1]{\lvert#1\rangle} 
\newcommand{\expect}[1]{ \langle #1 \rangle} 

\newcommand{\operator}[1]{\hat{#1}}

\newcommand{\mat}[1]{\mathsf{#1}}

\newcommand{\cone}{\mathrm{i}}

\renewcommand{\vec}[1]{\boldsymbol{#1}}

\begin{document}
\title{Ultrafast dynamics of electrons excited by femtosecond laser pulses: spin polarization and spin-polarized currents}

\author{Oliver Busch}
\email[Correspondence email address: ]{oliver.busch@physik.uni-halle.de}
\author{Franziska Ziolkowski}
\author{Ingrid Mertig}
\author{Jürgen Henk}
\affiliation{Institut für Physik, Martin Luther University Halle-Wittenberg, 06099 Halle, Germany}

\date{\today}

\begin{abstract}
Laser radiation incident on a ferromagnetic sample produces excited electrons and currents whose spin polarization must not be aligned with the magnetization---an effect due to spin-orbit coupling that is ubiquitous in spin- and angle-resolved photoemission. In this Paper, we report on a systematic investigation of the dynamics of spin polarization and spin-polarized currents produced by femtosecond laser pulses, modeled within our theoretical framework \textsc{evolve}. The spin polarization depends strongly on the properties of the laser pulse and on the sample composition, as is shown by comparing results for Cu(100), Co(100), and a Co/Cu heterostructure. We find a transition from coherence before the laser pulse's maximum to incoherence thereafter.  Moreover, the time dependence of the spin-polarization components induced by spin-orbit coupling differ significantly in Cu and Co: in Cu, we find long-period oscillations with tiny rapid modulations, whereas in Co prominent rapid oscillations with long period ones are superimposed. The pronounced spatial dependencies of the signals underline the importance of inhomogeneities, in particular magnetic/non-magnetic interfaces `act as source' for ultrafast spin-polarization effects. Our investigation provides detailed insight into electron dynamics during and shortly after a femtosecond laser excitation.
\end{abstract}


\maketitle

\section{Introduction}
Spin-polarized photocurrents are ubiquitous in spin- and angle-resolved photoelectron spectroscopy (SARPES)~\cite{Huefner1996,Schattke2003}. In non-magnetic samples, the spin polarization of the detected photocurrents---brought about by spin-orbit coupling (SOC)---depends on details of the setup, in particular on those of the incident electromagnetic radiation (e.g.\ on photon energy, polarization, and incidence direction; see for example Ref.~\onlinecite{Henk2018}) and on the symmetry of the surface~\cite{Tamura87,Tamura91a,Tamura91b,Henk1994}. In magnetic samples, the same effect results in magnetic dichroism~\cite{Henk1996,Starke2007}, and, as theoretical and experimental studies show, the spin-orbit-induced spin polarization \emph{of photoelectrons} must not be aligned with the magnetization direction (see for example Ref.~\onlinecite{Henk1996} and references therein).

In ultrafast spin dynamics, electrons are excited by electromagnetic radiation as well, for example by a femtosecond laser pulse. Focusing on the demagnetization of a magnetic sample~\cite{beaurepaire1996, zhang2000laser, zhang2003}, one investigates mainly the reduction of the magnetization but disregards its change in direction. The latter could be brought about by photo-induced spin-polarization components that are not aligned with the ground state's magnetization. In SARPES these `oblique' components are those of electrons measured \emph{at the detector}, whereas in ultrafast spin dynamics they are those of electrons \emph{within a sample}; thus one is concerned with different boundary conditions~\cite{Hermanson1977}. This idea immediately calls for a systematic investigation of photo-induced spin polarization and spin-polarized currents caused by femtosecond laser pulses~\cite{battiato2012theory}.  

In the theoretical study reported in this Paper, we concentrate on the SO-induced spin-polarization effects during a laser excitation. In order to excise the main features we begin with a non-magnetic sample---Cu(100)---and then turn to a magnetic sample---face-centered cubic Co(100). Since real samples often contain interfaces, we investigate the role of the latter by means of a Co/Cu(100) heterostructure. The simulations were performed using our computational framework \textsc{evolve}~\cite{Toepler2021}.

Questions worth considering are, among others, which components of the spin polarization are forbidden by symmetry? How large are the allowed components, and are their magnitudes comparable to those observed in SARPES? What are their temporal and spatial distributions? Does magnetism reduce the `oblique' SO-induced components (here: in samples containing Co)? What are the detailed properties of the photo-induced currents? We respond to these questions in this work.

This Paper is organized as follows. In Section~\ref{sec:models} we sketch our approach to ultrafast electron dynamics (\ref{sec:electron_dynamics}), discuss spin polarization as well as currents (\ref{sec:spinpol}), and perform a symmetry analysis (\ref{sec:symmetry}). Results are discussed in Section~\ref{sec:results}: beginning with Cu(100)~(\ref{sec:Cu}) we turn to magnetic systems, namely fcc Co(100) (\ref{sec:Co}) and a Co/Cu(100) heterostructure (\ref{sec:CoCu}). We conclude with Section~\ref{sec:conclusion}.

\section{Theoretical aspects}
\label{sec:models}
\subsection{Ultrafast electron dynamics}
\label{sec:electron_dynamics}
\begin{figure}
    \centering
    \includegraphics[width=0.99\columnwidth]{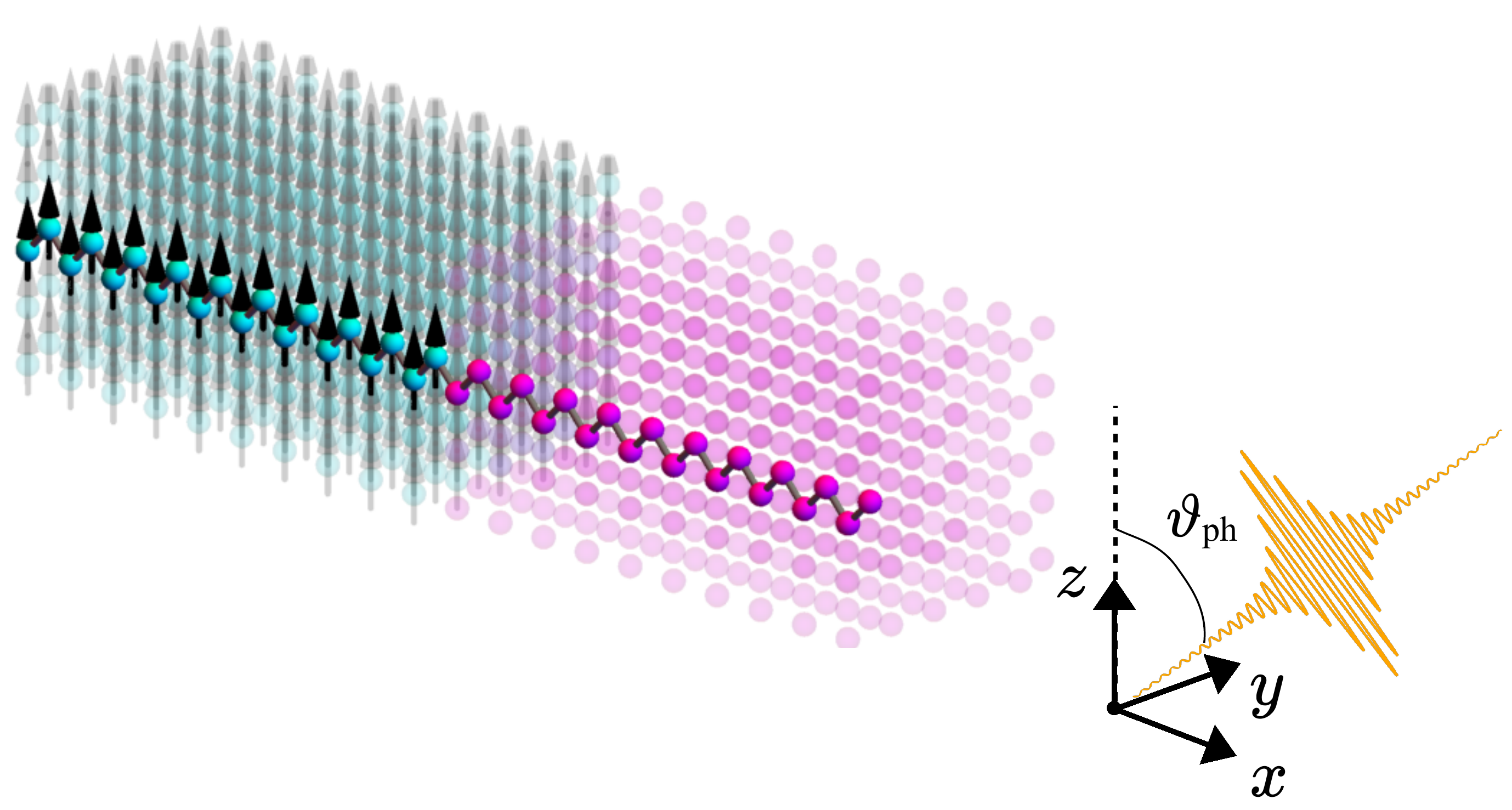}
    \caption{Geometry of a Co/Cu heterostructure. The fcc film consists of $40$~layers stacked in $x$ direction, with $20$~layers of Co atoms (cyan spheres) and $20$~layers of Cu atoms (magenta spheres). The Co magnetic moments point along the $z$ direction (black arrows). The film is infinite in $y$ and $z$ direction but finite in $x$ direction. A laser pulse impinges with a polar angle $\vartheta_{\mathrm{ph}}$ of $\unit[45]{^\circ}$ within the $xz$ plane onto the sample.}
    \label{fig:sketch_Co_Cu}
\end{figure}

The samples are free-standing fcc(100) films of $40$~layers thickness. We consider Cu(100), Co(100), and Co/Cu(100) (with $20$ layers each) films. The Cartesian $x$~axis is perpendicular to the film, and we apply periodic boundary conditions within the film, i.e.\ in~$y$ and~$z$~direction. In the case of Co(100) and Co/Cu(100), the magnetic moments are collinear and point along the $z$~direction (Fig.~\ref{fig:sketch_Co_Cu})~\cite{Heinrich91}.

The electronic structure of the samples is described by a tight-binding Hamiltonian~$\operator{H}_0$ of Slater-Koster type~\cite{Slater1954}, with parameters for the s-, p-, and d-orbitals taken from Ref.~\onlinecite{Papaconstantopoulos2015}. Collinear magnetism and spin-orbit coupling are taken into account as described in Ref.~\onlinecite{Konschuh2010}.

The electron system is excited by a femtosecond laser pulse with photon energy $E_{\mathrm{ph}} = \omega$  (in atomic units, $\hbar = 1$). The laser's electric field
\begin{align}
    \vec{E}(t) & = l(t) \sum_{l = \mathrm{s,p}} \vec{E}_{l} \cos(\omega t + \varphi_{l}).
    \label{eq:e-field}
\end{align}
is a coherent superposition of s- and p-polarized partial waves modulated with a Lorentzian envelope $l(t)$. $\vec{E}_{l}$ and $\varphi_{l}$ are the amplitudes resp.\ the phase shifts of the partial waves.

The electromagnetic radiation impinges within the $xz$~plane onto the films, with a polar angle $\vartheta_{\mathrm{ph}} = 45^{\circ}$ of incidence. For s-polarized light ($\vec{E}_{\mathrm{p}} = 0$), $\vec{E}(t)$ points along the $y$~axis which is perpendicular to the plane of incidence, the latter spanned by the incidence direction of the light and the surface normal. For p-polarized light ($\vec{E}_{\mathrm{s}} = 0$), $\vec{E}(t)$ lies within the $xz$ incidence plane.  Circular polarized radiation with helicity $\sigma^{\pm}$ is obtained by $\varphi_{\mathrm{s}} - \varphi_{\mathrm{p}} = \pm \unit[90]{^\circ}$ and equal amplitudes ($E_{\mathrm{s}} = E_{\mathrm{p}}$).

The electron dynamics is described by the von Neumann equation
\begin{align}
   -\cone \frac{\mathrm{d} \operator{\rho}(t)}{\mathrm{d} t} & =  [ \operator{\rho}(t),\operator{H}(t)]
    \label{eq:EOM}
\end{align}
for the one-particle density matrix
\begin{align}
    \operator{\rho}(t) = \sum_{n, m} \ket{n} \, p_{nm}(t) \, \bra{m}.
\end{align}
$\{ \ket{n} \}$ is the set of eigenstates of $\operator{H}_0$, with $\operator{H}_0 \ket{n} = \epsilon_{n} \ket{n}$. The time-dependent Hamiltonian $\operator{H}(t)$ comprises the electric field of the laser via minimal coupling~\cite{Savasta1995}. The equation of motion~\eqref{eq:EOM} for $\operator{\rho}(t)$ is solved within our theoretical framework \textsc{evolve}; for details see Ref.~\onlinecite{Toepler2021}. 

\subsection{Spin polarization and spin-polarized currents}
\label{sec:spinpol}
Site-, orbital-, and spin-resolved properties of an observable $O$ are obtained by taking partial traces in the expectation values $\expect{O}(t) = \operatorname{tr}[\operator{\rho}(t) \,\operator{O}]$, with the density matrix in an appropriate basis.

In matrix form an expectation value reads $\expect{O}(t) = \operatorname{tr}[\mat{P}(t) \,\mat{O}]$. We define matrices $\mat{p}_{kl}^{\sigma \sigma'}$ and $\mat{h}_{kl}^{\sigma \sigma'}$ for the density matrix and the Hamiltonian, respectively, with elements 
\begin{subequations}
\begin{align}
\left( \mat{p}_{kl}^{\sigma \sigma'} \right)_{\alpha \beta} & = p_{k \alpha \sigma, l \beta \sigma'},
\\
\left( \mat{h}_{kl}^{\sigma \sigma'} \right)_{\alpha \beta} & = h_{k \alpha \sigma, l \beta \sigma'}.
\end{align}    
\end{subequations}
$k$ and $l$ are site indices, $\sigma$ and $\sigma'$ specify the spin orientation ($\uparrow$ and $\downarrow$ with respect to the $z$ direction), and $\alpha$ and $\beta$ are orbital indices. These matrices are combined into site-resolved block matrices
\begin{subequations}
\begin{align}
\mat{P}_{kl} & = 
\begin{pmatrix}
\mat{p}_{kl}^{\uparrow \uparrow} & \mat{p}_{kl}^{\uparrow \downarrow} \\
\mat{p}_{kl}^{\downarrow \uparrow} & \mat{p}_{kl}^{\downarrow \downarrow}
\end{pmatrix},
\\
\mat{H}_{kl} & = 
\begin{pmatrix}
\mat{h}_{kl}^{\uparrow \uparrow} & \mat{h}_{kl}^{\uparrow \downarrow} \\
\mat{h}_{kl}^{\downarrow \uparrow} & \mat{h}_{kl}^{\downarrow \downarrow}
\end{pmatrix}.
\end{align}    
\end{subequations}

The spin polarization at site $l$ is given by
\begin{align*}
    s^{\mu}_{l} & = \operatorname{tr} \left( \mat{P}_{ll} \mat{\Sigma}^{\mu} \right), \quad \mu = x, y, z,
\end {align*}
in which $\mat{\Sigma}^{\mu}$ is a block Pauli matrix. Explicitly,
\begin{subequations}
\begin{align}
    s_{l}^{x} & = 2\, \mathrm{Re} \operatorname{tr} \left( \mat{p}_{ll}^{\uparrow \downarrow} \right), \\
    s_{l}^{y} & = -2\, \mathrm{Im} \operatorname{tr}\left( \mat{p}_{ll}^{\uparrow \downarrow} \right), \\
    s_{l}^{z} & = \operatorname{tr}\left( \mat{p}_{ll}^{\uparrow \uparrow} - \mat{p}_{ll}^{\downarrow \downarrow} \right),
\end{align}
\end{subequations}
with normalization $\operatorname{tr}(\mat{P}_{ll}) = 1$. The site-averaged spin polarization
\begin{align}
    S^{\mu} = \frac{1}{N_\mathrm{site}} \sum_{l} s_{l}^{\mu}, \quad \mu = x, y, z,
    \label{eq:spinpol-average}
\end{align}
is obtained by summation over all $N_\mathrm{site}$ sites in a film's unit cell.

The current
\begin{align}
j_{kl} & = 
- \frac{\mathrm{i}}{2} \operatorname{tr}
\left( \mat{P}_{lk} \mat{H}_{kl} \right) - \expect{l \leftrightarrow k}
\end{align}
from site $l$ to site $k$ as well as the respective spin-polarized currents 
\begin{align}
j_{kl}^{\mu} & = 
- \frac{\mathrm{i}}{4} \operatorname{tr}
\left( \mat{P}_{lk}
\left[ \mat{\Sigma}^{\mu}, \mat{H}_{kl} \right]_{+} \right)
- \expect{l \leftrightarrow k} \quad \mu = x, y, z,
\end{align}
are derived from Mahan's equation for the current operator in spin-symmetrized form~\cite{mahan2013} (see also Refs.~\onlinecite{Nikolic2006, Petrovic2018}; $\left[ \cdot, \cdot \right]_{+}$ is the anticommutator). For collinear magnetic textures, as discussed in this Paper, inter-site hopping with spin flip does not occur in $\operator{H}_{0}$, i.e. $\mat{h}_{kl}^{\uparrow \downarrow} = 0$ resp.\ $\mat{h}_{kl}^{\downarrow \uparrow}=0$. With this the above equations become
\begin{subequations}
\begin{align}
j_{kl} & = 
- \frac{\mathrm i}{2} \operatorname{tr} 
\left(
\mathsf{p}_{lk}^{\uparrow \uparrow} \mathsf{h}_{kl}^{\uparrow \uparrow} 
+
\mathsf{p}_{lk}^{\downarrow \downarrow} \mathsf{h}_{kl}^{\downarrow \downarrow} 
\right)
- \expect{l \leftrightarrow k}, \\
j_{kl}^{x} & = 
- \frac{\mathrm i}{4} \operatorname{tr}
\left( \mathsf{p}_{lk}^{\uparrow \downarrow} + \mathsf{p}_{lk}^{\downarrow \uparrow}\right)
\left( \mathsf{h}_{kl}^{\uparrow \uparrow} + \mathsf{h}_{kl}^{\downarrow \downarrow} \right) 
- \expect{l \leftrightarrow k}, \\
j_{kl}^{y} & = 
\frac{1}{4}\operatorname{tr}
\left( \mathsf{p}_{lk}^{\uparrow \downarrow} - \mathsf{p}_{lk}^{\downarrow \uparrow}\right)
\left( \mathsf{h}_{kl}^{\uparrow \uparrow} + \mathsf{h}_{kl}^{\downarrow \downarrow} \right) 
- \expect{l \leftrightarrow k}, \\
j_{kl}^{z} & = - \frac{\mathrm i}{2} \operatorname{tr} 
\left(
\mathsf{p}_{lk}^{\uparrow \uparrow} \mathsf{h}_{kl}^{\uparrow \uparrow} 
-
\mathsf{p}_{lk}^{\downarrow \downarrow} \mathsf{h}_{kl}^{\downarrow \downarrow} 
\right)
- \expect{l \leftrightarrow k}.
\end{align}
\end{subequations}
Interchanging the site and the spin indices yields $j_{kl} = - j_{lk}$ and $j_{kl}^{\mu} = - j_{lk}^{\mu} = j_{lk}^{-\mu}$.

\subsection{Symmetry analysis}
\label{sec:symmetry}
Instead of a full group-theoretical analysis~\cite{Henk1996}, we perform a symmetry analysis which tells what components of the spin polarization are forbidden for a given setup. The important symmetry is the reflection $\operator{m}_{y}$ at the $xz$ plane: $(x, y, z) \to (x, -y, z)$ since the $xz$ plane is a symmetry plane of the lattice and is also the laser's plane of incidence (spanned by the light incidence direction and the surface normal).

For p-polarized light, $\operator{m}_{y}$ is a symmetry operation for a non-magnetic sample ($\vec{M} = 0$; here: Cu) which tells that only $S^{y}$ is allowed nonzero (Table~\ref{tab:symmetry}). A $z$-magnetization breaks this symmetry ($\vec{M} \not= 0$; here: Co(100) and Co/Cu(100)), and all three components of $\vec{S}$ are allowed nonzero.

\begin{table}
	\caption{Effect of symmetry operations (left column) on the laser's electric field $\vec{E}$ decomposed into its s- and p-polarization components $\vec{E}_{\mathrm{s}}$ and $\vec{E}_{\mathrm{p}}$, the magnetization $\vec{M}$ in $z$ direction, and the electron spin polarization $\vec{S} = (S^{x}, S^{y}, S^{z})$. $\operator{1}$ is the identity operation, $\operator{m}_{y}$ is the reflection at the $xz$ plane.}
	\centering
	\begin{tabular}{c|rr|r|rrr}
		\hline \hline
		$\operator{1}$ &  $\vec{E}_{\mathrm{s}}$ &  $\vec{E}_{\mathrm{p}}$ & $\vec{M}$  &   $S^{x}$  &   $S^{y}$  &  $S^{z}$ \\
		$\operator{m}_{y}$ &  $-\vec{E}_{\mathrm{s}}$ &  $\vec{E}_{\mathrm{p}}$ &  $-\vec{M}$   &  $-S^{x}$ &   $S^{y}$  &  $-S^{z}$ \\
		\hline \hline 
	\end{tabular}
    \label{tab:symmetry}
\end{table}

For s-polarized light, the electric field of the laser is along the $y$ direction. Since for homogeneous non-magnetic samples (Cu) the $z$-rotation by $180^{\circ}$ leaves the setup invariant, $S^{y} = 0$ and $S^{z} = 0$. For $S^{y}$ this symmetry holds for the spin polarization at each site ($s_{l}^{y} = 0$). For $S^{z}$, however, it holds only for the site-averaged spin polarization, that is, $s_{l}^{z}$ at equivalent sites~$l$ may be nonzero but compensate each other (equivalent sites have the same distance from the two surfaces of a film).

Considering circular polarized light, $\operator{m}_{y}$ reverses the helicity $\sigma^{\pm} \to \sigma^{\mp}$ [$(\vec{E}_{\mathrm{s}}, \vec{E}_{\mathrm{p}}) \to (-\vec{E}_{\mathrm{s}}, \vec{E}_{\mathrm{p}})$], which tells that $S^{x}$ and $S^{z}$ change sign under helicity reversal for a non-magnetic sample but $S^{y}$ does not. For magnetic samples this strict relation is broken, which may be regarded as a magnetic spin dichroism (magnetic dichroism is an intensity change upon magnetization reversal~\cite{Ebert96}; here we are concerned with a change of the spin polarization). The symmetry-allowed or -forbidden spin polarization components are summarized in Table~\ref{tab:spinpol}.

\begin{table}
	\caption{Components of the site-averaged electron spin polarization $\vec{S} = (S^{x}, S^{y}, S^{z})$ allowed ($+$) or forbidden ($-$) by symmetry, for the magnetic case in rectangular brackets. For details see text.}
	\centering
 \begingroup
\setlength{\tabcolsep}{10pt} 
	\begin{tabular}{cccc}
		\hline \hline
	 polarization  & $S^{x}$  &   $S^{y}$  &  $S^{z}$ \\ 
  \hline
    p  &  $-$ [$+$] & $+$ [$+$] & $-$ [$+$] \\
    s  &  $-$ [$-$] & $-$ [$+$] & $-$ [$+$] \\
    circular & $+$ [$+$] & $+$ [$+$] & $+$ [$+$] \\
    \hline \hline 
	\end{tabular}
 \endgroup
    \label{tab:spinpol}
\end{table}

\section{Results and discussion}
\label{sec:results}
For discussing our results, we proceed by increasing step by step the order of complexity. We begin with a non-magnetic Cu(100) film, since it exhibits the phenomena most clearly. The effect of magnetism is addressed by fcc Co(100), and eventually the combination of both systems into a Co/Cu(100) heterostructure allows examining the effect of a magnetic/non-magnetic interface. 

In all simulations discussed below, the laser has a photon energy of~$\unit[1.55]{eV}$, a fluence of about $\unit[3.3]{mJ\,cm^{-2}}$, and is modulated with a Lorentzian $l(t)$ with $\unit[10]{fs}$ width. All samples comprise $40$~layers, with sites~$0$ and~$39$ defining the bottom and top surfaces, respectively.

\subsection{Cu(100)}
\label{sec:Cu}
\begin{figure}
    \centering
    \includegraphics[width=0.9\columnwidth]{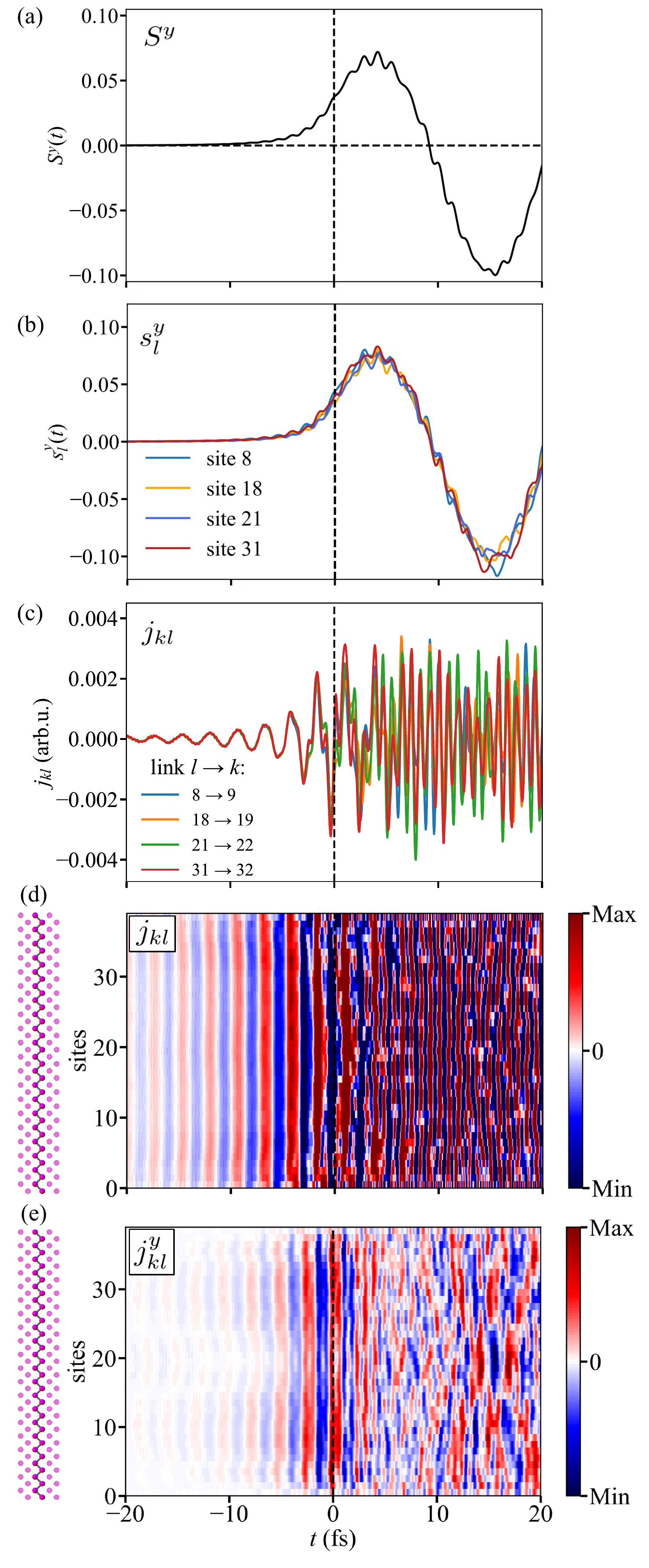}
    \caption{Photo-induced spin polarization and currents for a Cu(100) sample excited by p-polarized light. (a) Site-averaged spin polarization~$S^{y}(t)$. (b) Local spin polarization $s_{l}^{y}(t)$ for selected sites, as indicated. (c) Currents $j_{kl}(t)$ between neighboring sites $l \to k = l + 1$ for selected site pairs as indicated. (d) Currents $j_{kl}(t)$ and (e) spin-resolved currents $j_{kl}^{y}(t)$; their magnitude is indicated by color bars with the same range (red positive, blue negative). Data in panels~(c), (d) and (e) in arbitrary units. Vertical dashed lines at $t = \unit[0]{fs}$ mark the maximum of the laser pulse.}
    \label{fig:pure_Cu_p_pol}
\end{figure}

In accordance with the symmetry analysis (Table~\ref{tab:spinpol}), the calculations for p-polarized light yield only a nonzero $S^{y}$ that is slightly modulated with the doubled laser frequency (Fig.~\ref{fig:pure_Cu_p_pol}a). The sizable magnitude is explained by the local contributions~$s_{l}^{y}(t)$ which oscillate in phase with almost identical amplitude [constructive interference; panel~(b)]. After the laser pulse; deviations among the site-resolved spectra increase marginally (cf.\ $t > \unit[12]{fs}$). 

The above `unison' oscillations found for $s_{l}^{y}(t)$ show up as well in the currents $j_{kl}(t)$ before the laser pulse's maximum [Fig.~\ref{fig:pure_Cu_p_pol}(c) and (d)], but with a much smaller period. The laser's photon energy of $\unit[1.55]{eV}$ corresponds to a period of $\unit[2.7]{fs}$ or about $3.7$ oscillations within $\unit[10]{fs}$, which is also seen in panels~(c) and~(d). This suggests that the electron system follows the electric field of the laser, that is a collective motion across the film (in $x$ direction). At about $t = \unit[-3]{fs}$ increasing interference, starting at the surfaces, reduces the coherence in the oscillations, thereby obliterating the pattern at later times. 

The oscillations of the currents are accompanied by those of the spin-resolved currents $j_{kl}^{y}(t)$ in opposite direction [panel~(e); the $x$ and $z$ components are zero]. A current in positive $x$ direction (red in panel~d) appears simultaneously with a spin-polarized current in opposite direction [blue in panel~(e)], which implies a flow of $-y$-polarized electrons in $x$ direction. Again, the current pattern becomes complicated after the laser pulse due to the interferences mentioned before.

\begin{figure}
    \centering
    \includegraphics[width=0.9\columnwidth]{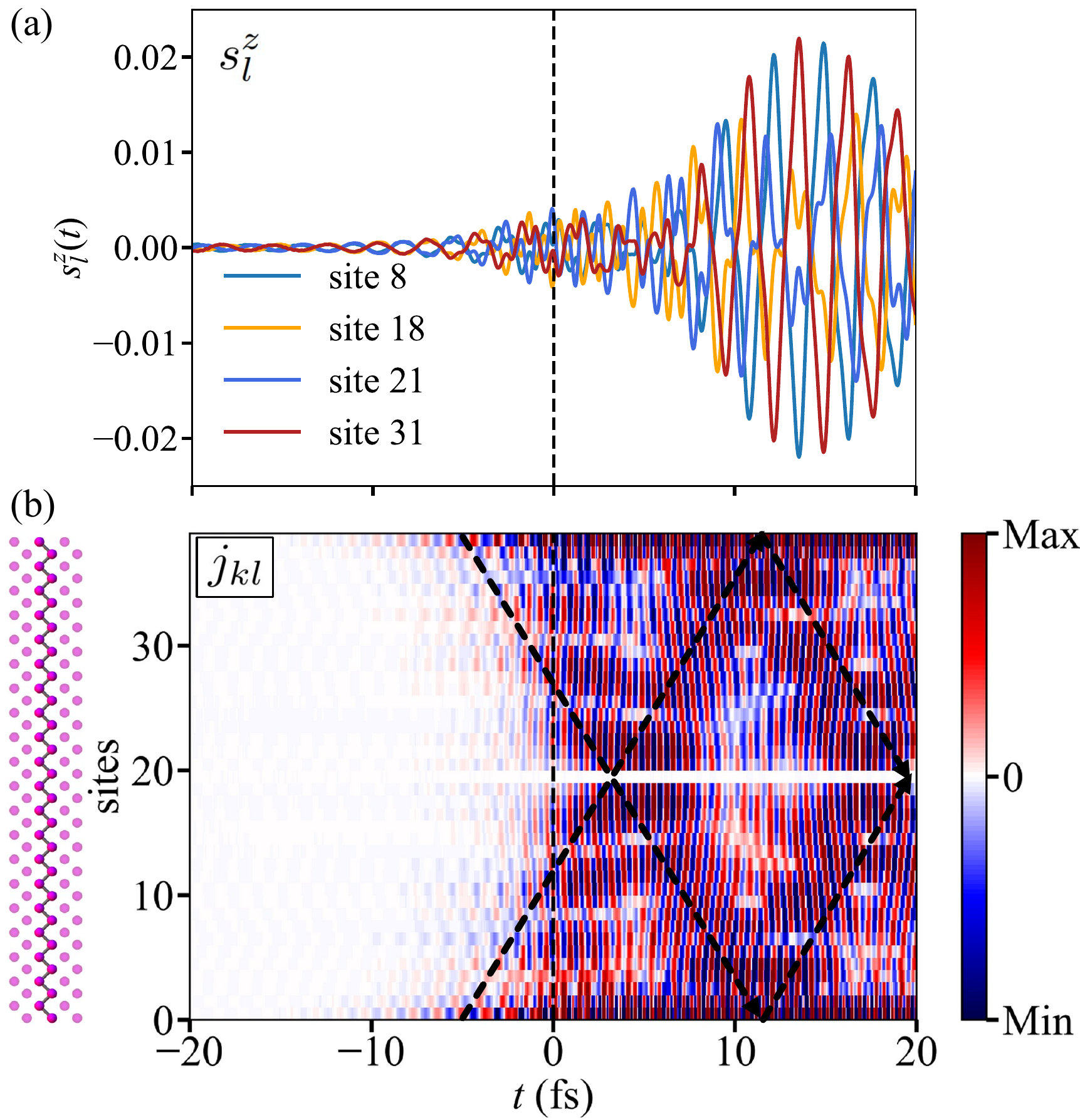}
    \caption{Photo-induced spin polarization and current of a Cu(100) film excited by s-polarized light. (a) $s_{l}^{z}(t)$ for selected sites as indicated. Sites $8$ ($18$) and $31$ ($21$) are equivalent. (b) Currents $j_{kl}(t)$ displayed as color scale (red positive, blue negative; in arbitrary units). Dashed arrows serve as guides to the eye. Vertical dashed lines at $t = \unit[0]{fs}$ indicate the maximum of the laser pulse.}
    \label{fig:pure_Cu_s_pol}
\end{figure}

For s-polarized light, the symmetry analysis yields $\vec{S} = 0$ but allows for $s^{z}_{l} \not= 0$. The photo-induced local spin polarizations at equivalent sites thus have to compensate each other. This is fully confirmed by the simulations: the spin polarization is spatially antisymmetric within the Cu film [Fig.~\ref{fig:pure_Cu_s_pol}(a)].

\begin{figure*}
    \centering
    \includegraphics[width=0.99\textwidth]{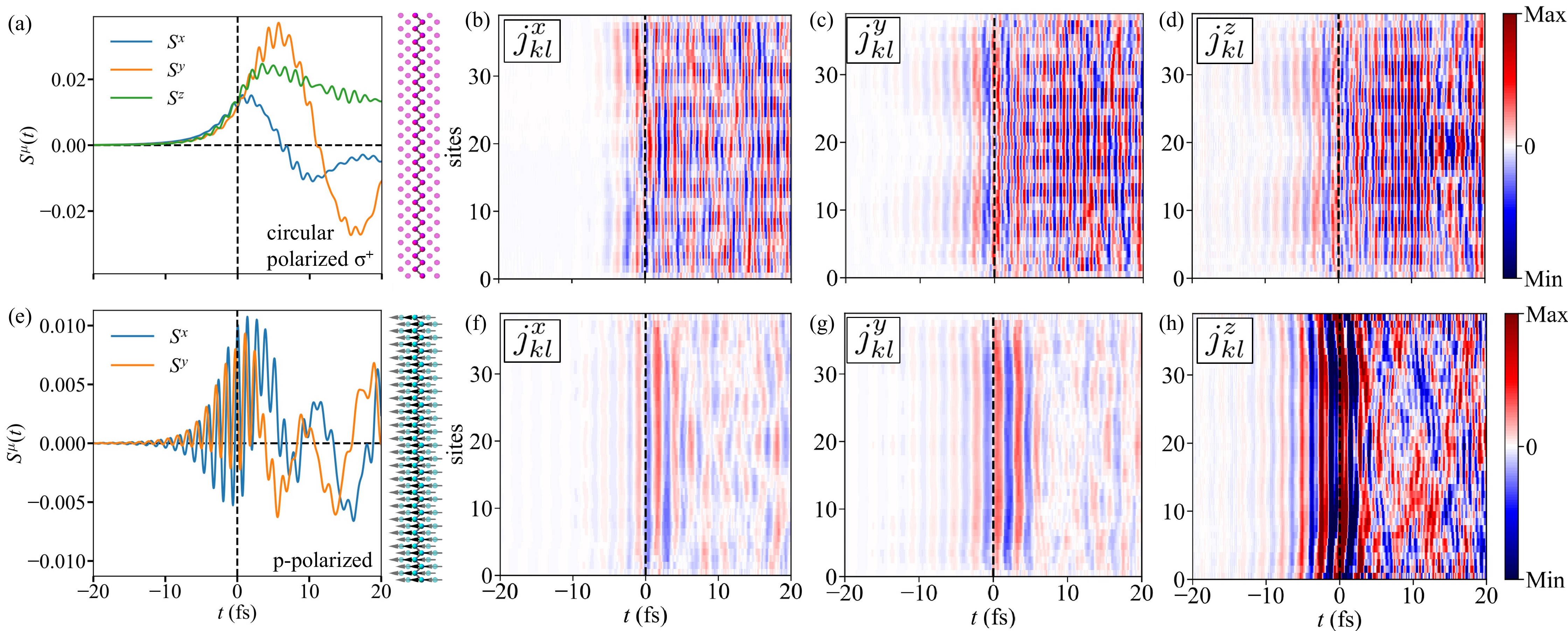}
    \caption{Photo-induced spin polarization $S^{\mu}(t)$ and spin-resolved currents $j_{kl}^{\mu}(t)$ for a Cu(100) film excited by circular polarized light with helicity $\sigma^{+}$ (top row) and for fcc Co(100) excited by p-polarized light (bottom row). (a) Site-averaged spin polarization $S^{\mu}(t)$ for Cu(100) ($\mu = x, y, z$). (b) -- (d) Spin-resolved currents $j_{kl}^{\mu}(t)$ displayed as color scale, as in Fig.~\ref{fig:pure_Cu_p_pol}. (e) $S^{x}(t)$ and $S^{y}(t)$ for Co(100). (f) -- (h) as (b) -- (d) using the same color scale. Dashed vertical lines indicate the maximum of the laser pulse at $t = \unit[0]{fs}$.}
    \label{fig:pure_Cu_circ_pol_pure_Co_p_pol}
\end{figure*}

The antisymmetry of the spin polarization may be attributed to the surface normals of the freestanding Cu film being opposite to each other. This reasoning complies with spin polarization effects in spin- and angle-resolved photoemission~\cite{Tamura87, Tamura91a, Tamura91b, Henk1994, Henk1996}, since these rely on the presence of a surface (they do not occur in bulk samples). Hence, one may regard the present result as a first indication for the importance of surfaces and interfaces for ultrafast spin dynamics; see for example Ref.~\onlinecite{Toepler2021} (for reviews on polarized electrons at surfaces we refer to Refs.~\onlinecite{Feder86} and~\onlinecite{Kirschner85}).

The above argument is supported by the currents $j_{kl}(t)$ [Fig.~\ref{fig:pure_Cu_s_pol}(b)] which are initiated at the surfaces: compare for example the darker color scale at the surface sites~0 and~39 in panel~(b) with respect to the lighter colors in the interior of the film at $t = -\unit[5]{fs}$. The currents enter the film's interior slightly after the laser's maximum (at $t \approx \unit[4]{fs}$), as schematically indicated by the dashed arrows (due to the antisymmetry, the current at the film's center vanishes, giving rise to the white horizontal stripe) and are reflected at the surfaces at $t \approx \unit[12]{fs}$, leading to a `crisscross' pattern [cf.\ the dashed arrows in panel~(b)]. The spin-resolved currents $j^{z}_{kl}(t)$ exhibit a pattern (not shown here) reminiscent of that of $j^{y}_{kl}(t)$ for p-polarized light displayed in Fig.~\ref{fig:pure_Cu_p_pol}(e). 

For circular polarized light it is sufficient to discuss one helicity (here: $\sigma^{+}$ as defined in Section~\ref{sec:electron_dynamics}), since the $x$ and $z$ components of both spin polarization and spin-resolved currents change sign upon helicity reversal, whereas the $y$ component does not, as is confirmed by our simulations.

All components of the site-averaged spin polarization $S^\mu(t)$ and the spin-resolved currents $j^\mu_{kl}(t)$ are nonzero (Fig.~\ref{fig:pure_Cu_circ_pol_pure_Co_p_pol}; top row). In an admittedly simple picture $S^{x}(t)$ and $S^{z}(t)$ may be viewed as due to optical orientation in photoemission~\cite{Kessler85}. Recall that the laser impinges within the $xz$ plane onto the film; for a single atom optical orientation by circular polarized light would then cause spin polarization within the $xz$ plane. Likewise, $S^{y}(t)$ may be attributed to the effect predicted by Tamura, Piepke, and Feder~\cite{Tamura87} for SARPES\@. Of course, this `decomposition of effects' ignores that the superposition of the laser's s- and p-polarized partial waves are coherent and shifted in phase. Moreover, the electron dynamics mixes the components of the local spin polarization because of spin-orbit coupling; nevertheless $S^{y}(t)$ is reminiscent of that for p-polarized light [Fig.~\ref{fig:pure_Cu_p_pol}(a)].

\subsection{fcc Co(100)}
\label{sec:Co}
For fcc Co(100) we focus on excitation by p-polarized light (bottom row in Fig.~\ref{fig:pure_Cu_circ_pol_pure_Co_p_pol}). As expected and often found in both experiment and theory, the site-averaged spin-polarization component $S^{z}(t)$ associated with magnetism is reduced by the laser pulse, that is the sample becomes demagnetized (cf.\ Ref.~\onlinecite{Ziolkowski23} and references therein). This demagnetization is site-dependent (not shown) similar to the induced spin polarization in Cu(100) discussed before. 

\begin{figure*}
    \centering
    \includegraphics[width=0.9\textwidth]{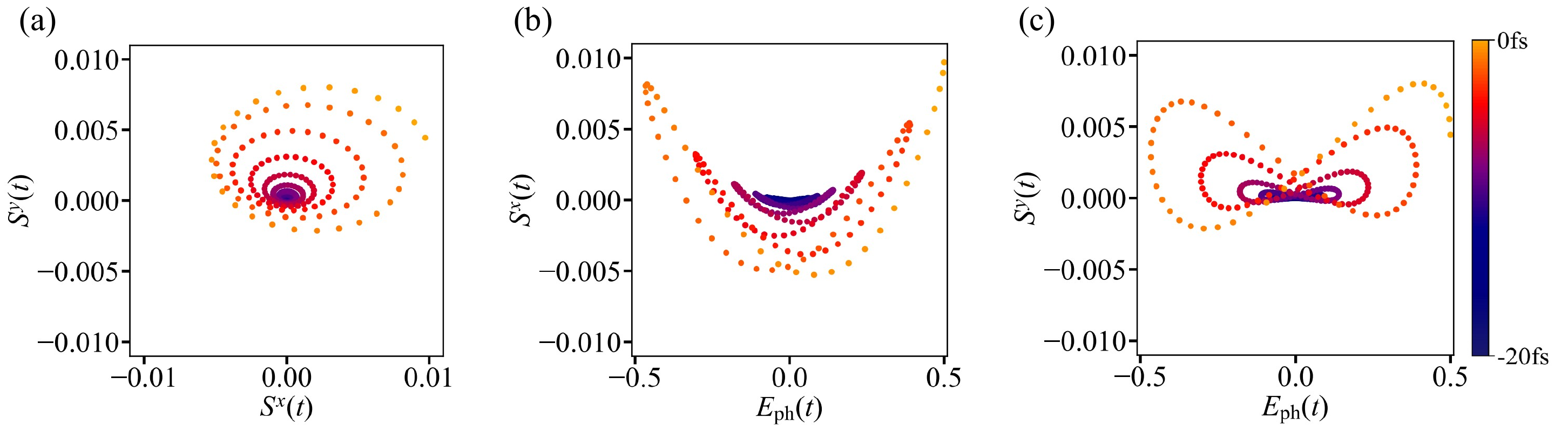}
    \caption{Laser-driven precession of the spin polarization in Co(100) excited by p-polarized light. The color scale visualizes the time evolution from $t = \unit[-20]{fs}$ (dark blue) to $t = \unit[0]{fs}$ (orange). (a) Correlation of $S^{y}(t)$ and $S^{x}(t)$ using data presented in Fig.~\ref{fig:pure_Cu_circ_pol_pure_Co_p_pol}e. Panels~(b) and~(c) show $S^{x}(t)$ resp.\ $S^{y}(t)$ versus the electric field $E_\mathrm{ph}(t)$ of the laser pulse.}
    \label{fig:pure_Co_precession}
\end{figure*}

In contrast to non-magnetic Cu(100), the magnetization of Co(100) breaks the mirror symmetry at the $xz$ plane and allows for nonzero $S^{x}(t)$ and $S^{y}(t)$; confer Table~\ref{tab:spinpol}. Both components are modulated by the doubled laser frequency but shifted in phase [panel~(e)]. Their magnitudes are roughly $\unit[10]{\%}$ of the $S^{y}$ component in Cu(100) [Fig.~\ref{fig:pure_Cu_p_pol}(a)]. Moreover, both $S^{x}(t)$ and $S^{y}(t)$ of Co(100) exhibit a beating pattern (with maxima at about $t \approx \unit[0]{fs}$, $ \unit[10]{fs}$ and $\unit[20]{fs}$), while $S^{y}(t)$ of Cu(100) displays a clear sinusoidal shape. 

The spin-polarization components $S^{x}(t)$ and $S^{y}(t)$ exhibit a regular pattern before the maximum of the laser pulse [Fig.~\ref{fig:pure_Cu_circ_pol_pure_Co_p_pol}(e)], which hints toward laser-driven precession of the spin polarization $\vec{S}(t)$. Indeed, $S^{x}(t)$ and $S^{y}(t)$ display a left-handed helix, starting at the origin and with increasing amplitude [Fig.~\ref{fig:pure_Co_precession}(a)]. Moreover, the noticeable shift of the spiral center to positive values is explained by spin-orbit coupling: a minimal tight-binding model for the motion of $\vec{S}(t)$, including SOC, yields two features: a deformation of the precession cone and a shift of the cone axis off the magnetization direction ($z$ axis). Without spin-orbit coupling, one finds the usual circular cone with its axis along the magnetization direction. 

The time sequences of $S^{x}(t)$ and $S^{y}(t)$ versus the laser amplitude $E_{\mathrm{ph}}(t)$ prove that the precession is driven by the laser [panels~(b) and~(c) in Fig.~\ref{fig:pure_Co_precession}]. The differences in the patterns are attributed to the phase shift between $S^{x}(t)$ and $S^{y}(t)$.

The striking differences in the spin polarization of Cu and Co could be attributed to the electronic structure, to spin-orbit coupling or to exchange splitting. Concerning the electronic structure, Cu has completely occupied d-orbitals, while the respective spin-resolved orbitals in Co are partially occupied. On the other hand, the electronic structure of Co might be viewed as that of Cu but exchange-split (in a rigid-band model). In this scenario, the differences could be ascribed to the magnetism in Co.

In order to shed light upon the origin we performed simulations for Cu and Co with varied strengths of the spin-orbit coupling and of the exchange splitting (not shown here). Varying the SOC strength has minute effect on the spin polarization in both Cu and Co. However, reducing the exchange splitting in Co(100) reveals that, first, the long-period oscillations in $S^{x}(t)$ and $S^{y}(t)$ are modified without clear trend, and second, the rapid oscillations associated with the laser pulse become suppressed. Without exchange splitting $S^{x}(t)$ vanishes and $S^{y}(t)$ displays features that are similar to those in Cu(100) [Fig.~\ref{fig:pure_Cu_p_pol}(a)]. These findings prompt magnetism as origin. 

As for the spin polarization, all three components of the spin-resolved currents are nonzero [panels~(f) -- (h) in Fig.~\ref{fig:pure_Cu_circ_pol_pure_Co_p_pol}], with the $z$-component $j_{kl}^{z}(t)$ being the largest [as exhibited by darker colors in panel~(h)]. All components oscillate `unison' before the laser pulse maximum; complicated current patterns arise after the pulse. 

Summarizing briefly for Cu and Co, we find that the simulations confirm the symmetry considerations. General trends are `unison' oscillations before the laser maximum and complicated patterns thereafter; the optically induced spin-polarization components are smaller in a magnetic sample but exhibit precession before the laser pulse maximum.

\subsection{Co/Cu heterostructure}
\label{sec:CoCu}
\begin{figure}
    \centering
    \includegraphics[width=0.9\columnwidth]{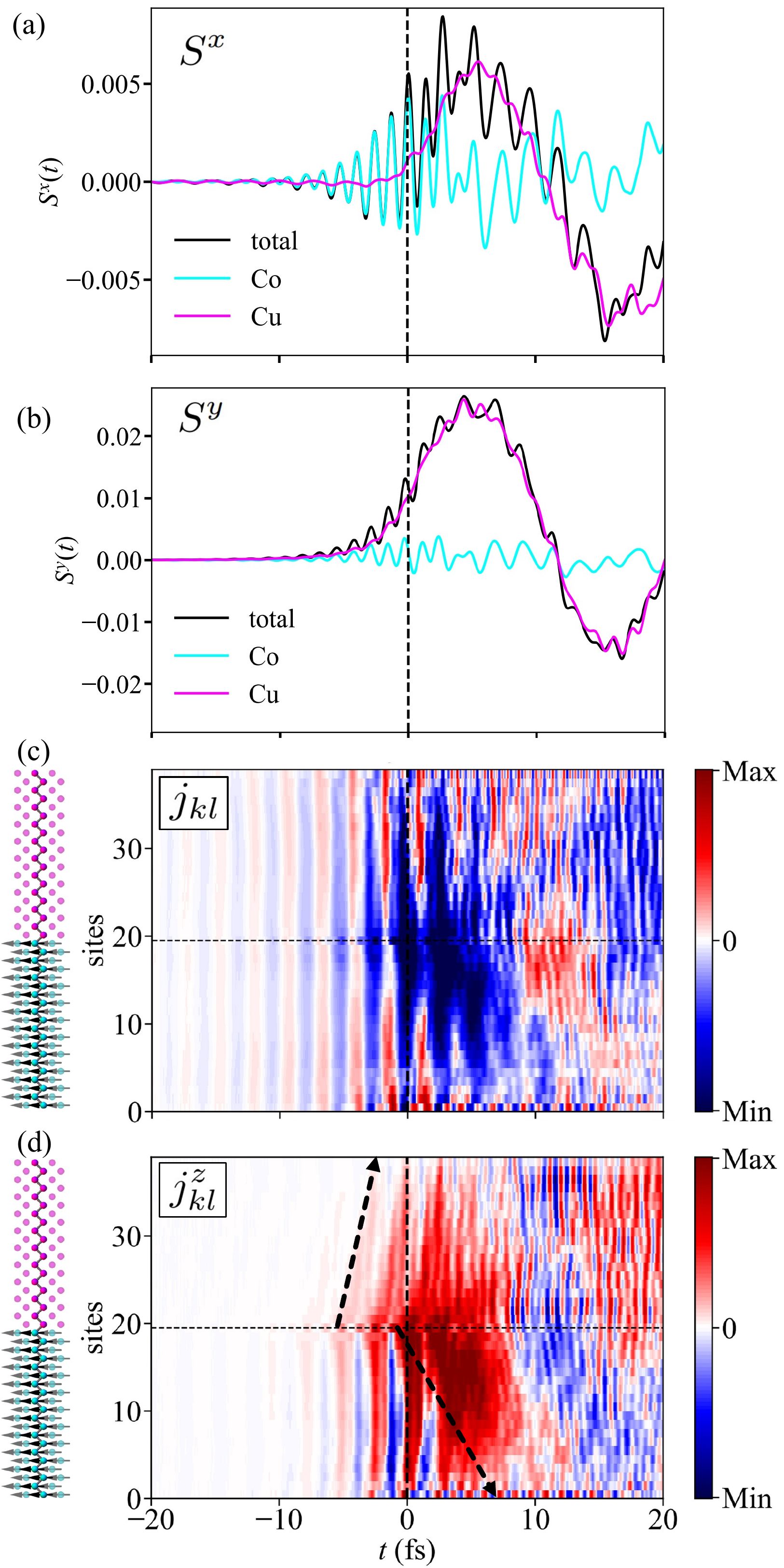}
    \caption{Photo-induced spin polarization and currents of a Co/Cu(100) heterostructure excited by p-polarized light. (a) Component $S^{x}(t)$ of the site-averaged spin polarization (black) decomposed into that in the Co region (cyan) and that in the Cu region (magenta). The latter are normalized with respect to $N_{\mathrm{site}}$; cf.\ Eq.~\eqref{eq:spinpol-average}. The maximum of the laser pulse at $t = \unit[0]{fs}$ is marked by the vertical dashed line. (b) As (a), but for $S^{y}(t)$. (c) Currents $j_{kl}(t)$ depicted as color scale (red positive, blue negative). The Co/Cu interface is identified by the horizontal dashed line. (d) As (c), but for spin-resolved currents $j_{kl}^{z}(t)$. Arrows serve as guide to the eye.}
    \label{fig:Co_Cu_p_pol}
\end{figure}

We now address a Co/Cu(100) heterostructure illuminated by p-polarized light. Decomposing $S^{x}(t)$ and $S^{y}(t)$ of the entire sample [black in Figs.~\ref{fig:Co_Cu_p_pol}(a) and (b)] into the respective parts in the Co (cyan) and in the Cu region (magenta) tells that $S^{x}(t)$ [panel~(a)] is first induced by the laser pulse in the Co region and enters subsequently the Cu region (recall that $S^{x}(t)$ is symmetry-forbidden in Cu(100); Section~\ref{sec:Cu}). This finding underlines the importance of an interface for ultrafast spin dynamics.  

In contrast, $S^{y}(t)$ is by far the largest in the non-magnetic Cu region [panel~(b)], whereas it is strongly reduced in the Co region. This finding corroborates the above argument that magnetism may reduce photo-induced spin-polarization components. Both the magnitude and frequency of the site-averaged components in the two regions are reminiscent of those in the respective homogeneous samples.

The currents $j_{kl}(t)$ exhibit an oscillating collective motion across the sample before the pulse, similar to Cu(100) [Fig.~\ref{fig:pure_Cu_p_pol}(c)]. However, beginning slightly before the pulse maximum at $t = \unit[0]{fs}$, the spatial homogeneity is lost; instead there are sizable currents initiated at the interface (visualized by the horizontal dashed line at site~19), and propagating towards the Co region [dark blue features in panel~(c)]. This finding corroborates that the interface acts as a `source' of ultrafast spin currents. At the magnetic/non-magnetic interface, the imbalance of occupation facilitates the production of currents. Moreover, since the imbalance is spin-dependent, also the spin-resolved currents $j_{kl}^{z}(t)$, that is those with spin along the magnetization direction, should be triggered at the interface. This is indeed verified by $j^{z}_{kl}(t)$ [panel~(d)]. More precisely, these currents are homogeneous in the Co region before the pulse; they become enhanced at the interface at about $t = \unit[-5]{fs}$ (dark red patches; also illustrated by the black arrows). The $x$- and $y$-spin-resolved currents (not shown here) are not as much affected by the interface as the $z$-component, which suggests that the imbalance of magnetization (spin-dependent occupation) at the interface is the most relevant origin.

The above argument concerning the importance of interfaces is further supported by the different velocities of $j_{kl}^{z}(t)$ in the Cu and in the Co region (cf.\ the slopes of the arrows). In the latter, we find the homogeneous oscillating current pattern before the pulse maximum. In the Cu region, which is non-magnetic, the same pattern appears oblique, as indicated by the black dashed arrow in panel~(d). This means that these currents `spill out' from the Co region into the Cu region and propagate toward the Cu surface (site~39).

\section{Conclusion and outlook}
\label{sec:conclusion}
Our theoretical findings suggest that femtosecond laser pulses impinging on thin films may be used for generating ultrafast oscillating spin-polarized currents. Moreover, interfaces amplify the production of these currents, as is evidenced in our study. And the spin polarization can be tuned by details of the laser's electric field, in particular by the polarization of the radiation.

Inhomogeneities in the sample (surfaces, interfaces) yield intrinsic imbalances of occupation which facilitate the production of spin-polarized currents. This finding supports reasoning given in Ref.~\onlinecite{Rouzegar22}, in which it is argued that a spin voltage, that is a spin-dependent imbalance of occupation, results in both demagnetization and spin currents. Moreover, the transfer of spin polarization from the magnetic Co region into the non-magnetic Cu region of a Co/Cu heterostructure may also be regarded as spin pumping: fluctuating magnetic moments in a ferromagnet produce spin currents in an attached normal metal \cite{tserkovnyak2002a,tserkovnyak2002b}. Hence, our study gives further details on the mechanisms for the transfer of spin polarization across a magnetic/non-magnetic interface due to laser excitation, as reported for example in Refs.~\onlinecite{Battiato2010, Battiato2013, Nenno2018, Alekhin2017}. 

As is shown in this Paper, already the combination of 3d materials (here: Co and Cu) produces sizable spin-polarization effects. The latter could be enhanced further by increasing the imbalance of spin-dependent occupation at interfaces. Material combinations worth investigating could comprise heavier elements with larger SOC (e.\,g.\ Pt) and heavy magnetic materials (e.\,g.\ Gd).

A direct observation of the photo-induced spin polarization and the spin-polarized currents studied in this Paper challenges experiments because of their limited temporal resolution. However, it is conceivable to probe the currents via their emitted electromagnetic radiation.
As a source of THz radiation  usually an electric dipole along $y$-direction is discussed, which is generated by a spin current $j^z$. In contrast a spin current $j^y$, carrying a spin polarization orthogonal to the interface normal and orthogonal to the magnetization direction generates an electric dipole along $z$-direction and could thus be distinguished from the former one experimentally.  The emitted radiation can be computed from the density matrix and the currents using the Jefimenko equations~\cite{Jefimenkow1966,Ridley2021} and thus be compared with measured signals (e.\,g.\ Ref.~\onlinecite{Kampfrath2013}).

\acknowledgements
This work is funded by the Deutsche Forschungsgemeinschaft (DFG, German Research Foundation) -- Project-ID 328545488 -- TRR~227, project~B04.

\bibliographystyle{apsrev4-2}
\bibliography{references}

\end{document}